# Tuning inter-dot tunnel coupling of an etched graphene double quantum dot by adjacent metal gates

Da Wei[1], Hai-Ou Li[1], Gang Cao[1], Gang Luo[1], Zhi-Xiong Zheng[1], Tao Tu[1], Ming Xiao[1], Guang-Can Guo[1], Hong-Wen Jiang[2] & Guo-Ping Guo[1]

**Abstract**

Graphene double quantum open the possibility to use charge or spin degrees of freedom for storing and manipulating quantum information in this new electronic material. However, impurities and edge disorders in etched graphene nano-structures hinder the ability to control the inter-dot tunnel coupling, $t_C$, the most important property of the artificial molecule. Here we report measurements of $t_C$ in an all-metal-gates-tuned graphene DQD. We find that $t_C$ can be controlled continuously about a factor of four by employing a single gate. Furthermore, $t_C$, can be changed monotonically about another factor of four as electrons are gate-pumped into the dot one by one. The results suggest that the strength of tunnel coupling in etched graphene DQDs can be varied in a rather broad range and in a controllable manner, which improves the outlook to use graphene as a base material for qubit applications.

**Introduction**

Graphene, a newly discovered material which is fabricated from a few layers of its bulky origin, graphite, has attracted much recent attention for its promising variety of nano-electronics applications. Among these novel applications, several proposals have been made to store and manipulate quantum information in graphene based quantum dots, attracted by its possible long coherence time, unique band structures, and ability to oppose localization et al [1, 2].

For quantum dot based quantum information processing, double quantum dots (DQDs), sometimes called artificial molecules, emerge as one of the leading candidates for qubit implementation, which can be encoded either by the bonding and anti-bonding states of charges or by the singlet and triplet states of spins. In order to perform logic operations in a DQD, for either charge qubits or spin qubits, it is essential to measure and to gain control over the tunnel coupling between two adjacent dots.  This tunneling coupling is also fundamentally important for two-qubit operations. Indeed, such control has been demonstrated in DQDs made of more conventional semiconducting materials, such as GaAs/AlGaAs [3] and Si/SiGe [4]. So far graphene DQD have already been successfully fabricated by a number of groups.[5-8] Although in an earlier transport measurement of a more conventional graphene DQD [6], $t_C$ was qualitatively estimated from the curvature of honeycomb edge patterns, suggesting the inter-dot coupling can be very strong, controllable inter-dot tunneling coupling in etched graphene DQDs remains to be a challenge. This may derive from the fact that the etched graphene DQD system suffers from unpredictable and usually non-monotonic changes in its inter-dot configuration. When one tries to tune the system by adjusting the local electrostatic potential, mainly due to impurities and edge states' capturing/releasing electrons, such reconfiguration happens and results in non-monotonic changes in inter-dot capacitances.[9-12] The difficulty of experimentally controlling tunneling barriers in graphene have commonly lead to an intuitive thinking that it is nearly impossible to control the $t_c$ coupling by tuning the gate voltages applied to the QD proximity.

[1]Key Laboratory of Quantum Information, Department of Optics and Optical Engineering, University of Science and Technology of China, Chinese Academy of Science, Hefei 230026, China. [2]Department of Physics and Astronomy, University of California at Los Angeles, California 90095, USA. Correspondence and requests for materials should be addressed to G.P.G. (email: gpguo@ustc.edu.cn)

Here, we demonstrate that our experimental studies may suggest otherwise. In our all-metal-gates-tuned graphene DQD. The actual electron temperature ($T_e$) under base temperature is accurately determined by using a refined measurement method. As a consequence, both capacitative coupling of the DQD charge network and the inter-dot tunneling coupling can be accurately measured. Our results show that the inter-dot tunnel coupling $t_c$ strength can be tuned about a factor of 4 simply by tuning the inter-dot potential barrier through gates. Such tunability appears to be exponentially monotonic as expected for a tunneling barrier defined DQD system. In addition, we also find the relative electron number in one dot can varied to change $t_c$ by another factor of 4, as one adds electrons into one quantum dot one by one. Such tendencies appear to be monotonic over as many as nine consecutive electrons.

Based on our experimental findings, we suggest that, contrary to conventional wisdom, etched graphene DQD systems, with multiple metal control gates, may still be an excellent platform for the fundamental research in this material aiming to demonstrate qubit operations. It is also reasonable to expect that with the rapid progress in nanotechnology, specifically with gaining more control over the impurity and the edge disorder in graphene ribbon structures, graphene DQDs may eventually become a viable basic building block for solid-state qubits.

**RESULTS**

**Electrostatic-gate controlled graphene double quantum dots.**

The device used for the experiment is a double quantum dot with an integrated charge state sensor. The base graphene structures of the DQD along with the adjacent detection channel are defined by plasma etching of large flake. The electrostatic control is facilitated by incorporating additional metallic gates, as shown in the SEM picture, in Fig. 1a. We have studied over ten identical devices. For consistency, the bulk of the data presented in this paper is from one sample.

Conventionally, the terminals of controlling gates for QDs are also formed by graphene ribbon structures.[6-12]  As experiments have shown that small graphene ribbon structure (i.e., smaller than 100nm) cannot be considered as a good conductor, as localized states are likely populated over the ribbon, we argue nano-ribbon an undesirable candidate for stable gating. Therefore to minimize the non-monotonic or sudden change in the electrostatic environment, in our device all controlling gates are made out of high-conductance metal using precision alignment for placing them in the close proximity, as close as 40nm, of the etched graphene ribbon structures.

As shown in Fig 1a. , the sample consists of a double-dot structure accompanied by a Quantum Point Contact (QPC) read-out channel down under. The QPC channel serves as a non-invasive detector of the charge states in the DQD system. Each dot has roughly a physical dimension of 120nm by 80nm, the connecting ribbon structure is of 100nm length and 35nm width. Specifically, the gates LP (RP) and LB (RB) are designed to control the electrochemical potential of the left dot (right dot) and left barrier (right barrier) respectively. Middle gate (marked as MG) is primarily used to tune the capacitive and the tunneling coupling of the two adjacent dots and gate Q is for the adjustment of the QPC's working point for optimum sensitivity.

**Determination of electron temperature of the device.**

The experiment is carried out in an Oxford Instruments top-loading dilution refrigerator. The sample is in direct thermal contact with the He3/He4 liquid mixture. The base temperature of the bath is 35mK and the bath temperature can be accurately varied from base to 700 mK.

Sweeping LP and RP at the same time we observed the characteristic signature of the formation of the DQD, known as the honeycomb diagram, as displayed for a typical sample in Fig1b. The charge stability diagram shows rather ordered structures for extended range of gate voltages. From such diagram we may extract nearly all the capacitances that form the charge network. Within each honeycomb, electron number for each dot is a well-defined integer. Thus, the combs are marked by (M+j, N+i) to denote the relative electron numbers in left and right dots, respectively.

Such diagrams are obtained through QPC serving as an electrostatic detector. It is possible to detect all the charge transition boundaries because the conductance of QPC channel depends sensitively on the electrochemical potential of the coupled DQD systems. For DQDs in Coulomb blockade regime, the electrons are only allowed to enter the dot one by one. The discreteness in charge brings a set of discrete electrochemical energy levels in each dot. Thus, entrance of an electron in one dot will cause a sudden change in QPC conductance.

As the QPC signal is an ensemble-averaged response to the DQD system, thermal fluctuation and quantum processes such as tunneling will smear the sharpness in the conductance shift. DiCarlo et.al [3] is the first to employ the sharpness of the conductance change as a function of the detuning, for an inter-dot transition, to investigate the effective electron temperature and the tunneling coupling strength of two adjacent dots. In our measurement, we refine their technique by using a modulation technique. A small modulation voltage is superimposed onto the DC scanning voltage on RP. The signal, detected by a lock-in amplifier, is the physical derivative of the original one, known as transconductance, dI/dV$_{RP}$, as the transconductance signal appears to have far better signal-noise ratio than the origin conductance signal. It turns out, for our graphene devices, due to the relatively large noise background, this refinement is very much necessary in order to reliably extract the temperature and the tunneling information.

As a consequence of charge transfer from one dot to another, the transconductance gives rise to a peak. The Full Width at Half Maximum in unit of detuning energy of such transition peak, denoted as FWHM (eV), directly reflects the sharpness of the corresponding conductance shift. The relation between FWHM (eV), $T_e$ and $t_C$ is [14]:

$$\frac{\Delta^2}{\sqrt{\Delta^2+F^2}^3} th(\sqrt{\Delta^2+F^2}) + \frac{F^2}{\Delta^2+F^2} ch^{-2}(\sqrt{\Delta^2+F^2}) = \frac{th\Delta}{2\Delta};$$

$$F = \frac{FWHM(eV)}{2} \cdot \frac{1}{2k_B T_e}; \qquad (1)$$

$$\Delta = \frac{t_C}{2k_B T_e}.$$

From this expression, it is clear that the FWHM is a combination of the two intertwining parameters of $T_e$ and $t_C$. It is therefore of critical importance to determine the electron temperature, which can be significantly different from the bath temperature, before a reliable $t_C$ value can be measured.

To determine $T_e$ we strategically select a place where $t_C$ is very small (i.e., $t_C \ll 2k_B T_e$). In this limit, the peak width of the transition line is then directly proportional to $T_e$. By Setting $t_C$ in eq.(1) to zero, one obtains the relation:

$$FWHM(eV) = 2 \cdot ln(\sqrt{2}+1) \cdot 2k_B T_e \approx 3.5 k_B T_e \qquad (2)$$

According to eq.(2), we investigate the evolution of peak width against raising mixing chamber temperature. Fig. 2 is obtained in the region where we find the tunneling strength weak enough to be neglected. We measure the peak width along the detuning line (white dashed line) as shown in Fig. 2a. Fig. 2b shows an example of this peak broadening for three mixing chamber temperatures. The results in Fig. 2c clearly display a linear dependence of FWHM on the mixing chamber temperature when temperature is high enough to provide a strong enough electron-phonon coupling so that electron is in equilibrium with the graphene lattice. From Fig. 2c, we extract the lever arm of the left plunger gate, $C_{gL}/C_L$=0.039. Taking into consideration that the $t_C$ of this measured transition line is no larger than $2\mu eV$ (0.5GHz), which is estimated through the intercept of the dashed line in Fig. 2c, we finally calculate $T_e$ under dilution refrigerator's base temperature to be 75.3±9.4mK. This $T_e$ is verified for several transition lines, all in the weak-coupling region.

**Measurement and tunability of the inter-dot tunnel coupling strength.**

Once $T_e$ is obtained, we are now capable of measuring the $t_C$ coupling strength throughout the entire $V_{LP}$-$V_{RP}$ plane. For any particular transition line, we sweep $V_{LP}$ and $V_{RP}$ along the detuning line, as shown before, and obtain its peak width projected in either $V_{LP}$ or $V_{RP}$ axis. Take $V_{LP}$ for instance, the measured FWHM in unit of voltage, FWHM (V), is firstly converted into unit of detuning, FWHM (eV), then non-dimensionized using $F = FWHM(eV)/4k_BT_e$. After F obtained, solving eq.(1) about $\Delta$, eventually we can extract the absolute value of $t_C$ by multiply $\Delta$ with $2k_BT_e$.

Starting from the region showed in Fig. 3a, we have extracted $t_C$ for 13 inter-dot transition lines, marked by the fixed number of electrons for the right dot from N+8 to N-5. [15]

As shown in Fig. 3b, it is apparent that the $t_C$ increases monotonically as the number of electrons in the right QD is added one by one. The measured maximum value is 300ueV, for the "transition N+8", which corresponds a tunneling frequency of about 70 GHz and is more than 4 times the value for the minimum, for the "transition N". Such regularity over large number of charge configurations appears to against the common perception of that the tuning in etched QD is nearly uncontrollable. We speculate our device design of all-metal control gates may have improved the stability of the electrostatic environment to a level that is suitable for other coherent control experiments over a large space of gate voltages. It is important to mention that although regularity is demonstrated here over the entire plane, such monotonicity does not survive over very large range of charge configurations. As shown in the Fig. 3c, a sudden change has been observed. The possible mechanism of this anomalous behavior is likely due to the charge impurities in the substrate or edge localized states in the graphene nano-ribbon itself and will be discussed later.

In addition, we have studied the tunability of using a single metal gate. Fig. 4 shows the results for two inter-dot transition lines, namely, transition N+4 and N+5. As we vary the voltage on the middle gate (MG), both curves change rapidly and saturated at large voltages. The $t_C$ has been encouragingly changed by as much as a fact of 4 for a single gate. This voltage range is only about two unit of charging energy Ec. The change of $t_C$ can even be larger if we vary the voltages on other gates. Overall, with different gate combinations, we are able to vary the $t_C$ from 2 ueV to 400 ueV. This tunability in gate voltage also encouragingly suggests that such devices can be an excellent platform for the fundamental research in this material that aims at demonstration of qubit operations. We would like to point out the controllable range of $t_c$ in this work is well within the practical range for both coherent charge qubit [17,18] and spin qubit operations [19,20].

**DISCUSSION:**

Despite the encouraging facts that $t_c$ can be well-tuned either by changing the number of electrons in a dot or by varying the voltage of a metallic gate, for a given charge configuration, anomalous behaviors are also observed.

As shown in Fig. 3c, while t_c decreases monotonically as the number of electrons is reduced from N+8 to N, it abruptly increases from N-1 to N-5. The same thing happens when the tuning gate exceeds a certain voltage range, that is, when MG voltage is more negative than -150mV (~1.5$E_C$), the exponential increase suffers from a drop (not shown). We speculate that such anomalous behaviors are most-likely due to the edge-state-induced disorder, as they are shown to be inevitable in etched graphene ribbon structures [9-12]. The disorder can cause local electrochemical potential fluctuates and form randomly distributed puddle states, that may or may not contribute to the DQD transport. The metallic gate can change the electrostatic potential smoothly in a short range. However, over large voltage range the capacitive coupling of the metallic gates to the puddle states can also add or subtract charges discretely to these parasitic dots, thus altered the entire environment abruptly and unexpectedly. Consequently, the DQD needs to re-configure itself leading to changes in both dot-gate capacitances and interdot tunneling strength. To describe this conjecture more intuitively, we use the schematic illustration in Fig. 5 as an example of a possible scenario. The DQD wavefunctions in Fig. 5a are distorted as an electron is added discretely into one of the puddle states, as shown in Fig. 5b, which leads to an abrupt change in inter-dot coupling. More systematic investigation of the disorders in graphene nano-ribbons may shed light on the microscopic details of the DQD re-configuration and thus is called for.

However, the effects of the imperfections should not impede the research progress in the etched graphene DQD structure, as our result encouragingly show that the gate control of $t_c$ can be extended over a broad range of gate voltages and charge configurations. Of course, for graphene DQDs to become a viable basic building block for solid-state quibits, advancement in graphne processing nanotechnology is needed. With steady improvement in the quality of graphene nano-ribbon, we fully expect the $t_c$ tunability can eventually reach a level that is as good as that for DQDs made out of more conventional semiconductor, such as GaAs/AlGaAs [3] and Si/SiGe [4].

## METHOD:

**Device Fabrication.**

The devices used for the experiment are fabricated as follow. We mechanically exfoliate graphene sheets from a KISH graphite and located them using prefabricated metal marks. We then use electron-beam lithography (EBL) to define the nano-ribbon structures of the DQD as well as the charge detection channel. The unwanted portion is then etched by Inductively Coupled Plasma (ICP) using a mixture of $O_2$ and Ar (4:1). A second round of EBL is done using the precision alignment markers for placing Ti-Au (3nm/20nm) metal gates on top of the nano-ribbon structures. It is worth to note here, unlike the conventional fabrication techniques, we remove all the graphene materials other than the small DQD and QPC base structures. All the control gates are formed by metals instead of graphene to assure that gates are free of localized states.

**Converting the detuning value from voltage to energy.**

Signal is obtained when sweeping LP and RP voltage along the white dashed line as shown in Fig. 2a and Fig. 3a, thus the preliminarily obtained FWHM of the transition peak is in unit of LP (or RP) gate voltage. For consistency we project such width to LP axis and denote such FWHM as FWHM (V). Further calculation asks for a conversion from voltage to energy (detuning), therefore we denote the converted FWHM as FWHM (eV).

These two values are related by a set of capacitances of the charge network [16]:

$$FWHM(eV) = FWHM(V) \cdot \left[\frac{C_{gL}(C_R-C_M)}{C_L C_R - C_M^2}\right] \cdot (1+k^2) \quad (3)$$

Here $C_{gL}$ is the capacitance between left plunger gate and the left dot. $C_L$ and $C_R$ are the total capacitance of the left and right dot respectively. $C_M$ is the interdot coupling capacitance. Last but not least, k is the slope of the detuning line, as the dashed white line shown in Fig 2a.

**Errors in the $t_C$ determination.**

The error bars indicated in Fig. 3 and Fig. 4 can be categorized into three parts: 1) the uncertainty in measured base electron temperature (~8%), 2) error from the lever-arm for each inter-dot transition (~7%) and 3) measurement error of all the capacitances used to convert the FWHM from voltage to energy (3%). Altogether, those errors lead to an average uncertainty of 15%.

Admittedly, the amplitude of the modulation stimulus will induce a broadening effect (~5%) to the measured peak, when such amplitude exceeds 40% of the FWHM

(V). However here in our measurements, we have minimized such effect by measuring each FWHM (V) under a set of different modulation amplitude to make sure that we've reached the best signal-noise ratio and the least broadening effect at the same time.

**Acknowledgements**

This work was supported by the National Fundamental Research Program (Grant No. 2011CBA00200), NNSF (Grant Nos. 11222438, 10934006, 11274294, 11074243, 11174267 and 91121014), and CAS. The authors thank M.L. Zhang, G.W. Deng and S.Y. Wang for the discussions and help.

**Author contributions**

D.W. fabricated the samples. D.W., G.L., H.O.L., G.C., M.X., G.P.G. and H.W.J. performed the measurements. D.W., Z.X.Z., T.T., and G.C.G. provided theoretical support and analyzed the data. The manuscript is prepared by D.W., H.W.J. and G.P.G., G.P.G supervised the project. All authors contributed in discussing the results and commented on the manuscript.

**Additional information**

Supplementary information accompanies this paper is attached.

**Competing financial interests**

The authors declare no competing financial interests.

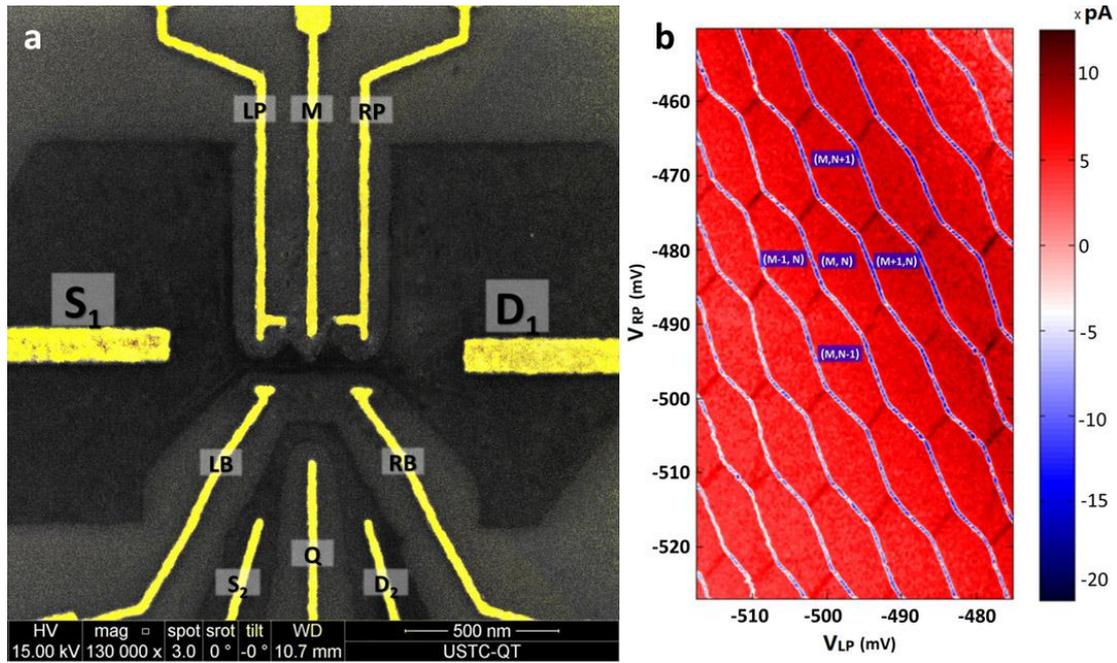

Fig. 1
  a. Scanning electron micrograph in false color of the device.   Dark regions are the graphene base structure consisting of a double quantum dot (100nm*100nm for each dot, 35nm*100nm for each ribbon) and an integrated QPC channel serving as in situ charge detector. Grey area is the etched-away region that shows the $SiO_2$ substrate surface. The yellow regions are the metal gates used for control and as source/drain electrodes.

  b. A typical charge sensor stability diagram of the double quantum dot: differential charge sensor current $dI/dV_{RP}$ as a function of two gate voltages $V_{LP}$ and $V_{RP}$. The structure shows a well-defined double quantum dot with characteristic honeycomb patterns. Charge state is defined by (M,N), where M and N are the electron number in the left and dot, respectively.[13]

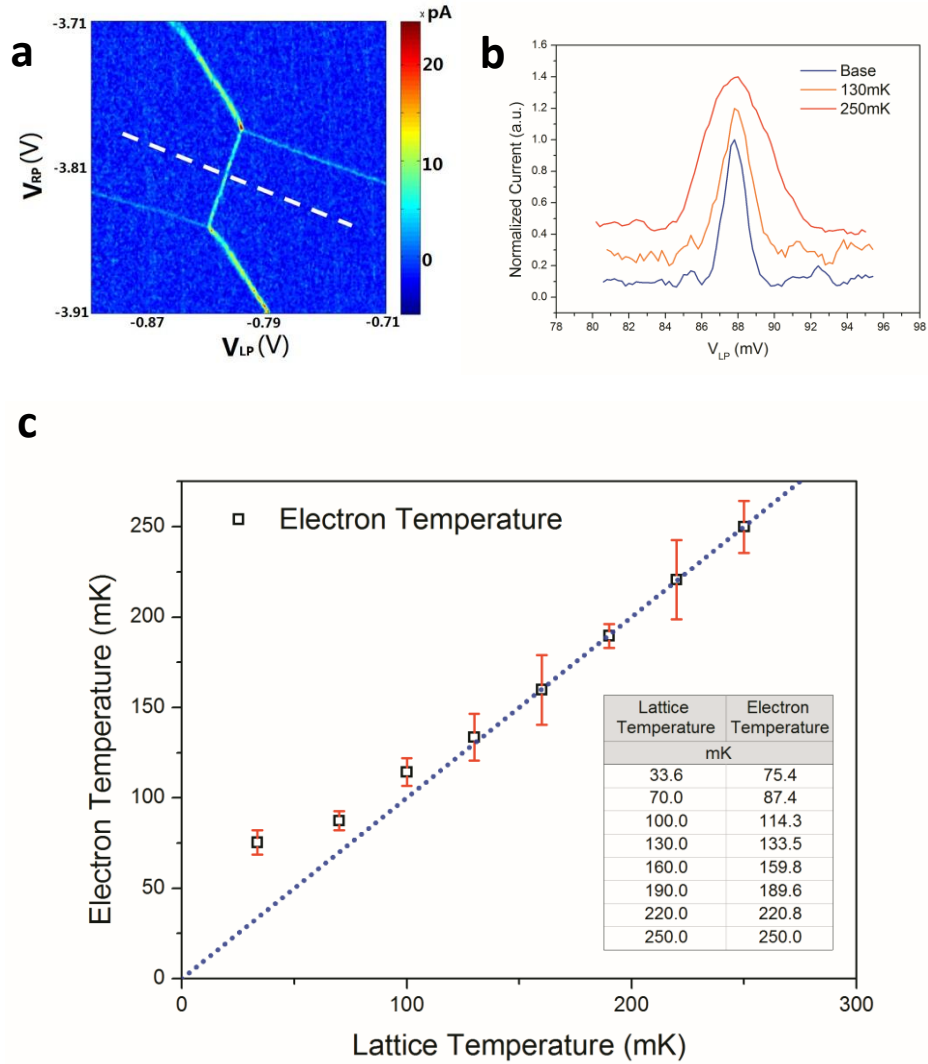

Fig. 2

a. Charge stability diagram in a weakly coupled region. The white dashed line indicates the scanning direction for the transition peak broadening study, which is perpendicular to the transition line and is along the detuning direction.

b. The curves are taken at different temperatures showing the temperature broadening effect of the transition peak. Curves are first normalized and then offset 0.2 for clarity.

c. Electron temperature, deduced from the peak-width, as a function of the lattice temperature. Dashed line is the best fit of the linear regime and its intercept reveals information of the coupling strength, from which we assure that $t_c$ here is no more than 2 μeV. Such calibration measures the electron temperature accurately and we eventually obtain that $T_e$ to be 75.3±9.4mK at the base temperature of the mixing chamber.

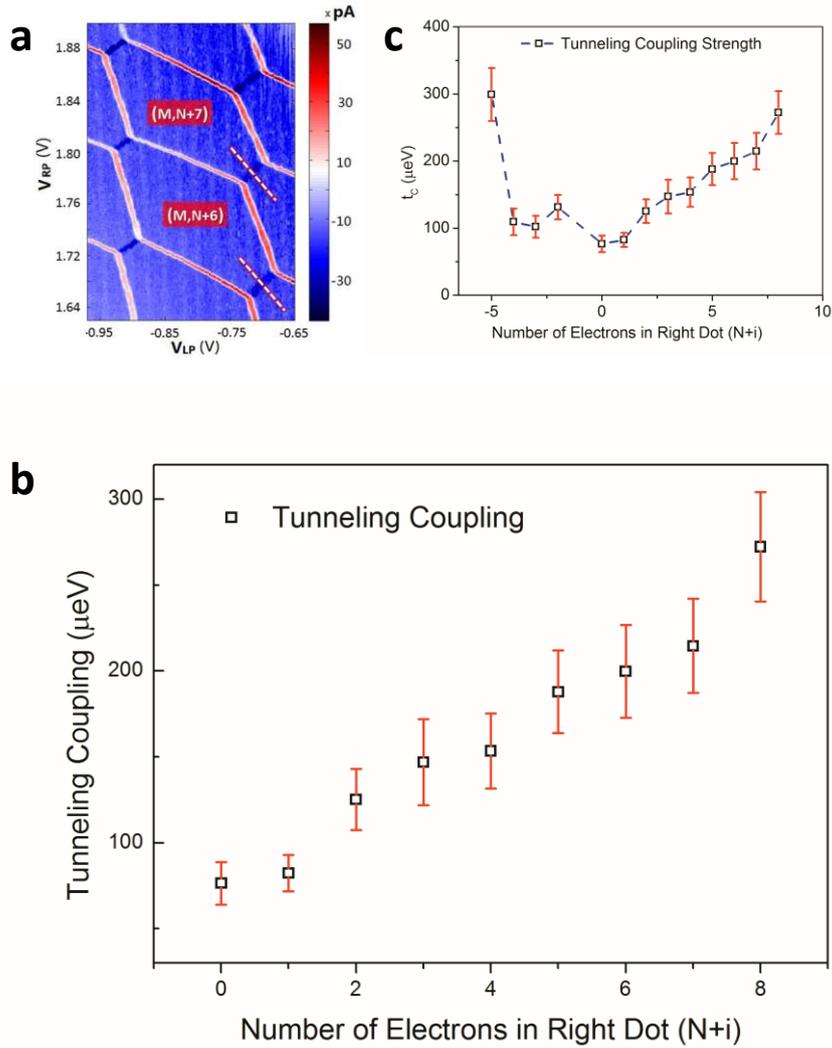

Fig. 3

a. Region where we start to investigate $t_c$'s dependence on electron number. Inter-dot tunnel coupling is measured along white dashed line, which involve with different number of electrons in the dots, as shown.[14]
b. $t_c$ undergoes a monotonic increase from 70 µeV to 300 µeV as electron number is increased.
c. When the measurement is extended over larger number of electrons, a sudden change is observed. The anomalous behavior is interpreted as a result of disorder in the graphene nano-ribbons, as discussed in the text and in Fig. 5.

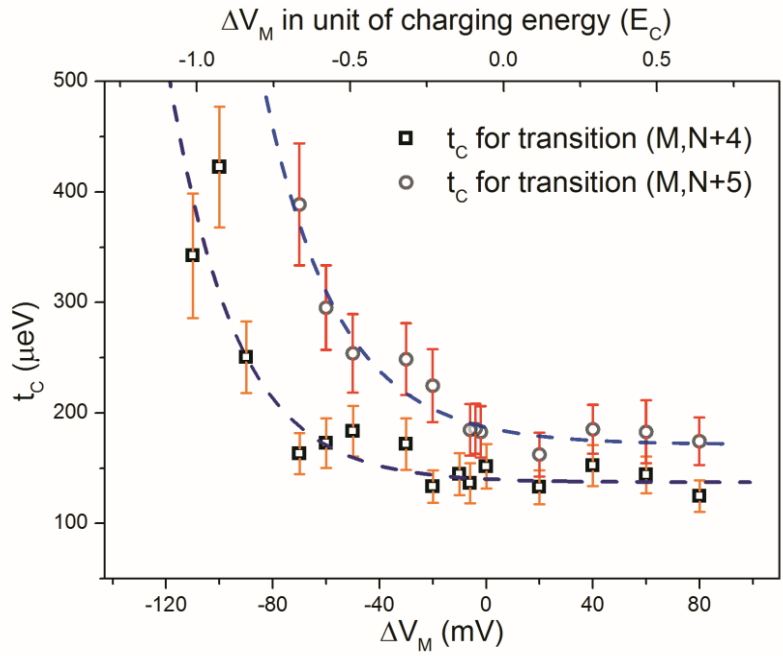

Fig. 4

Inter-dot tunnel coupling as a function of the middle gate voltage.  It shows that $t_C$ can be modulated by about a factor of 4 by applying gate voltage on a single gate.

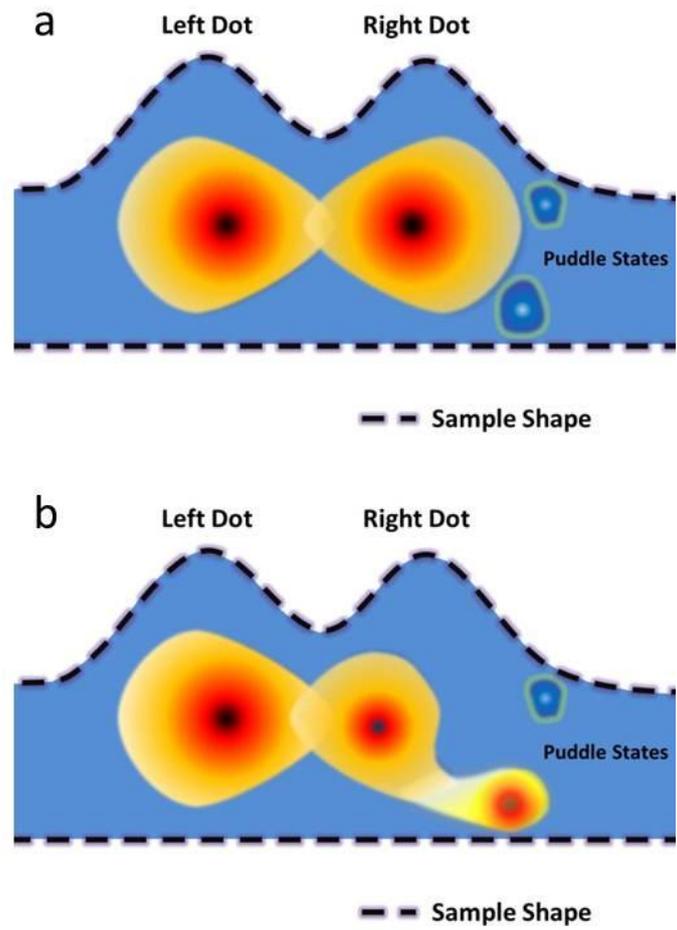

Fig. 5

Graphic illustration of an effect of puddle states on the inter-dot tunnel coupling. The DQD wavefunctions in (a) are distorted leading to an abrupt change in inter-dot coupling. (b), as an electron is added discretely into one of the puddle states.